# Milliwatt average power, MHz-repetition rate, broadband THz generation in organic crystal BNA with diamond substrate

SAMIRA MANSOURZADEH[1], TIM VOGEL[1], MOSTAFA SHALABY[2], FRANK WULF[1] AND CLARA J. SARACENO[1]

[1]*Photonics and Ultrafast Laser Science, Ruhr-Universität Bochum, Universitätsstrasse 150, 44801 Bochum, Germany*
[2]*Swiss Terahertz Research-Zurich, Technopark, 8005 Zurich, Switzerland and Park Innovaare, 5234 Villigen, Switzerland*
*\*Mansourzadeh.Samira@ruhr-uni-bochum.de*

**Abstract:** We demonstrate a 13.3 MHz repetition rate, broadband THz source with milliwatt-average power, obtained by collinear optical rectification of a high-power Yb-doped thin-disk laser in the organic crystal BNA (N-benzyl-2-methyl-4-nitroaniline). Our source reaches a maximum THz average power of 0.95 mW with an optical-to-THz efficiency of $4\times10^{-4}$ and a spectral bandwidth spanning up to 6 THz at -50 dB, driven by 2.4 W average power (after an optical chopper with duty cycle of 10%), 85 fs-pulses. This high average power excitation was possible without damaging the crystal by using a diamond-heatsinked crystal with significantly improved thermal properties. To the best of our knowledge, this result represents the highest THz average power reported so far using the commercially available organic crystal BNA, showing the potential of these crystals for high average power, high repetition rate femtosecond excitation. The combination of high power, high dynamic range, high repetition rate and broadband spectrum makes the demonstrated THz source highly attractive to improve various time-domain spectroscopy applications. Furthermore, we present a first exploration of the thermal behavior of BNA in this excitation regime, showing that thermal effects are the main limitation in average power scaling in these crystals.

## 1. Introduction

Terahertz time domain spectroscopy (THz-TDS) has progressed into a well-established tool in various fields in science and technology, such as the security sector for hidden object detection [1], biomedical research [2], or in fundamental research applications such as nanoscale imaging in condensed matter [3].

Among various techniques suitable to generate the required broadband, phase-stable THz pulses for TDS, optical rectification (OR) of ultrashort pulses in $\chi^{(2)}$ nonlinear crystals is one of the most commonly used, and a large variety of materials have been explored to improve THz source performance [4]. One area that has seen significant interest in the last years is the generation of high-repetition rate, high average power THz pulses for TDS using modern Yb-based high-power driving ultrafast lasers, for improving signal-to-noise-ratio and/or measurement times. In this respect, Gallium Phosphide (GaP) is a popular choice, because it offers broadband operation (typically beyond 6 THz) and velocity matching is achieved in a simple collinear scheme for 1030 nm, where modern Yb-based driving lasers typically operate [5–7]. Among a variety of these novel high-average power laser systems that are becoming increasingly available [8–10], we focus our attention on mode-locked thin-disk oscillators, as they provide pulses in the MHz repetition rate regime at high average power, directly from a one-box oscillator with similar output power levels as complex multi-chain amplifiers [11]. Using this technology, we recently demonstrated mW-average power THz levels with bandwidths up to 6 THz using collinear OR in GaP, excited by a 112 W laser system [7]. However, even in this moderate pulse energy regime, these semiconductor materials suffer from intrinsically moderate nonlinear coefficients and from two-photon absorption when pumping at 1 µm [12], which limit the conversion efficiency to typical values <$10^{-5}$. To circumvent these limitations, the tilted-pulse front technique in Lithium Niobate (LiNbO$_3$) is typically applied, at the expense of the simplicity of the setup [13]. Using this technique, average powers in the 100 mW range [14,15] have been recently achieved. However, the emitted THz spectral bandwidth is confined to a frequency range below 2 THz due to phonon absorption in Lithium Niobate, and it was recently shown that conversion efficiencies become limited in this geometry when using µJ pump pulse energies available at MHz repetition rate [16].

Organic crystals are an attractive alternative to combine wide bandwidth with high conversion efficiency up to the percent-level in a collinear scheme. However, so far, it was commonly accepted that high excitation average powers were not possible to apply at moderate pulse energies (i.e. with small spot sizes) in these crystals, due to their poor thermal properties; therefore, most results using organic crystals were so far restricted to very low repetition rates <100 Hz with high excitation pulse energies (in the multi-mJ regime) [17,18]. Only very recently, the organic crystal HMQ-TMS was investigated at MHz repetition rate which resulted in mW level THz average power [19]. In this experiment, a compressed Yb-doped fiber laser with a peak power of ~9 MW at 10 MHz was used to achieve 1.38 mW of THz average power. The spectrum extended up to 6 THz at -30 dB. Despite this result being very promising, HMQ-TMS remains a very specialized crystal with low availability, and higher excitation average powers were not possible to apply. In this respect the commercially available BNA [20] is a promising alternative, which has been extensively investigated at lower repetition rates. Using a 800 nm pump wavelength and an adjustable repetition rate between 10 Hz to 100 Hz [21], an optical-to-THz conversion efficiency of 0.2% with peak to peak electric field of ~4 MV/cm has been obtained. In [22], a 1150-1550 nm near infrared pump at moderate repetition rate of 1 kHz was used to achieve a high optical-to-THz conversion efficiency of 0.8% and record strong fields of 10 MV/cm. However, the THz generation properties of BNA crystal and its potential to generate high average power THz radiation in MHz repetition rate range at 1030 nm driving wavelength were so far still unknown. In particular, thermal effects and damage threshold at high-repetition rate (and correspondingly small spot sizes) were never explored in the goal of understanding scaling laws.

Here, we demonstrate THz generation in BNA, driven by an ultrafast laser with 13.3-MHz repetition rate and 85-fs pulse duration at 1030 nm central wavelength. We use bursts of the driving laser pulses with variable duty cycle, to optimize efficiency and THz power, as well as gain understanding of the limiting factors. Under optimized conditions, we obtain 0.95 mW of THz average power with a conversion efficiency of $4\times10^{-4}$, by heatsinking the crystal to a diamond substrate. To the best of our knowledge, this is the highest THz average power obtained from BNA. Moreover, we perform a systematic investigation of THz generation as a function of average power and pulse energy as well as the thermal load on the BNA crystal, showing that by better thermal management of the generation crystal, further upscaling to the several tens of milliwatt seems possible. This will open the door for broadband THz sources combining high repetition rate and high average power.

## 2. Experimental setup

The full experimental setup is shown in Fig .1. The driving laser is a home-built semiconductor saturable absorber mirror mode-locked Yb-doped thin-disk oscillator delivering up to 110 W at 13.3 MHz repetition rate with a central wavelength of 1030 nm and a pulse duration of 570 fs, resulting in a maximum pulse energy of 9.2 µJ. A Herriott-type multi-pass cell (MPC) [23] can be used to spectrally-broaden the laser output and reduce the pulse duration to 85 fs via chirp compensation, with a high transmission of 96%, resulting in 106 W of average power with this short pulse duration. More details about this laser system have been presented in [7]. After the MPC, the laser beam is guided towards the THz-TDS setup and then split in two parts: 99% of the total power is available to be used as pump for THz generation and around 1% of the total power is used as the probe beam for electro-optic sampling (EOS, see below). To adjust the laser power in the generation arm, a combination of a waveplate ($\lambda/2$) and a thin film polarizer (TFP) is used. Note, that in order to avoid crystal damage, in this study we only use up to 30 W of the maximum available laser power. To control the thermal load, an optical chopper is placed before the BNA crystal. The 10-slot chopper blade has an adjustable slot opening, allowing us to vary the duty cycle between 10%-50%. This results in pump pulse bursts with variable duration and burst energy. Consequently, we can keep either the used pump energy or the average power constant while varying the other quantity. This allows us to disentangle the influence of average power, thermal effects, and peak intensity on these crystals.

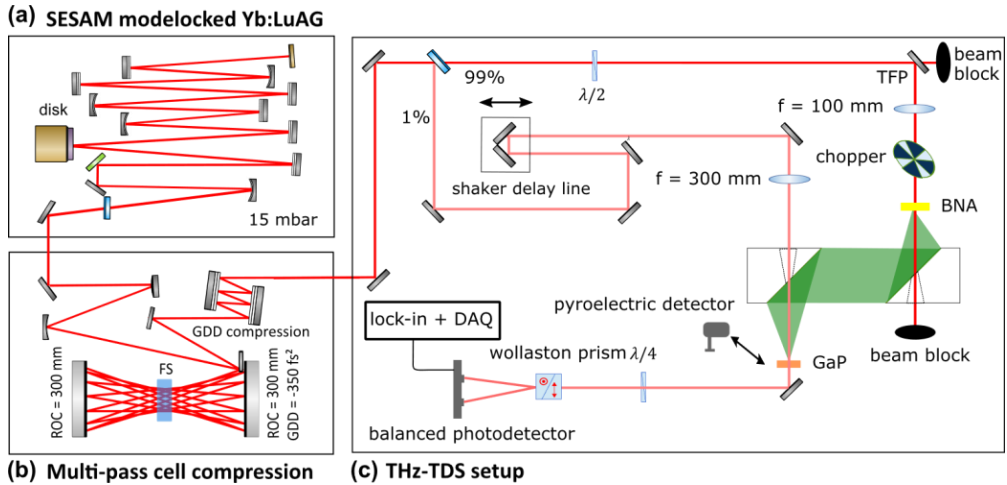

Fig. 1. Full experimental setup consisting of (a) mode-locked thin disk laser, (b) Herriot type MPC and compression mirrors and 106 W average power are available after the MPC, (c) THz-TDS setup. Using a combination of the TFP and $\lambda/2$, the power on the BNA is controlled and the residual power goes to a beam block.

A commercial BNA crystal (supplied by Swiss Terahertz GmbH) is used as a generation crystal. It is placed in the focus of a 100-mm focal length convex lens to focus the pump beam on the position of the crystal down to 0.25 mm ($1/e^2$ diameter) in all experiments. The generated THz radiation is collected and refocused on the detector using two 2" off-axis parabolic (OAP) mirrors with focal lengths of 50 mm and 100 mm, respectively. We characterize the THz radiation using either a calibrated pyroelectric power sensor (THz 20, SLT Sensor- und Lasertechnik GmbH) or a standard electro-optic sampling (EOS) setup. The EOS consists of a sampling crystal, a $\lambda/4$ plate, followed by a Wollaston prism and a balanced photodetector. A anti reflection (AR)-coated <110>-cut GaP crystal with a thickness of either 0.2 mm or 0.5 mm is used as the sampling crystal and was placed in the focus of the second OAP. The delay between pump and probe beam is provided by an oscillating delay line. An additional chopper with a frequency of 18 Hz is used in combination with the power sensor. Additionally, we blocked off any residual pump light with a set of filters including a Polytetrafluoroethylene (PTFE) tape (with 89% THz transmission) and black textile (with 30% THz transmission). The data is acquired using a lock-in amplifier (Zurich Instrument, UHFLI) in combination with a data acquisition system (Dewesoft, SIRIUS M) which records the demodulated signal out of the lock-in amplifier and the digitized position of the shaker. The modulation frequency of the pump beam required as a reference for the lock-in and is set to 963 Hz, the maximum possible frequency provided by the adjustable chopper blade is 1 kHz. The bandwidth of 120 Hz is chosen for the low-pass filter of the lock-in and the frequency of the shaker to sample the THz trace is 0.55 Hz.

To investigate the thermal behavior of the THz generation in BNA, we measure the temperature of the BNA crystal in all experiments with an infrared camera (InfraTec, VarioCam HD). To obtain an accurate temperature measurement, we first characterize the emissivity of the BNA crystal. In this goal, the crystal is placed on a heat plate with a matte black aluminum surface underneath. The temperature of the heat plate is varied from 30°C to 60°C in 5°C steps and at each point, the temperature of the crystal as well as the black surface as a reference are measured by the camera which looks to the surface of the BNA. By considering the ambient temperature of 20°C, the calculated emissivity at each temperature is shown in Table 1. The average emissivity value of BNA in the beforementioned temperature range and under an angle of 20 degrees is 0.87 which we used to estimate the real temperature of the crystal in the following sections.

Table 1. Emissivity measurement of BNA crystal on diamond substrate measured at 20°C of environmental temperature.

| Setpoint /°C | 30 | 35 | 40 | 45 | 50 | 55 | 60 |
|---|---|---|---|---|---|---|---|
| Measured temperature of BNA with $\varepsilon = 1$ /°C | 27.19 | 30.83 | 34.63 | 38.51 | 42.27 | 46.29 | 49.86 |
| Calculated emissivity | 0.89 | 0.88 | 0.87 | 0.87 | 0.87 | 0.87 | 0.86 |

## 3. Results

In order to investigate the THz generation properties of the BNA crystal and characterize its thermal behavior, four experiments are presented. First, the pulse energy on the crystal is kept constant and the pump power is varied, and we measure the THz electric field and the average power for different duty cycles at fixed energy. Then, we keep the average power on the crystal constant and vary the pulse energy by changing the duty cycle of the chopper. Afterwards, to achieve the highest average power value using BNA, we measure THz power vs. pump power for an optimized duty cycle. For each experiment, we additionally monitor the crystal temperature with an infrared camera.

### 3.1 THz generation efficiency vs. duty cycle at constant pulse energy on the crystal

In this section, the pump pulse energy is kept constant on the crystal and the pump average power is changed by varying the duty cycle of the chopper. In each duty cycle of 10% to 50%, pump pulse bursts with variable duration are generated which results in different average power on the crystal after the chopper. The effective repetition rate (number of the pulses per second) varies from 6.7 MHz to 1.3 MHz when the duty cycle reduces from 50% to 10%. It should be mentioned that in order to keep the pulse energy on the crystal constant, the waveplate before the chopper stays in a fixed rotation angle during the duty cycle variation. The pulse energy on the crystal is 0.14 µJ but the average power is increased from 0.18 W to 0.9 W when the duty cycle increases from 10% to 50%, respectively. Fig. 2 shows schematically the pump pulse bursts at constant pulse energy and variable average power (red dashed lines) by changing the duty cycle. The bursts with duty cycle of 50% and 20% are shown in Fig. 2a and Fig. 2b, respectively. The dashed red line indicates the relative average power which decreases by reducing the duty cycle, although the amplitude of each single pulse (proportional to the pulse energy) is constant by varying the duty cycle.

The BNA used in this section and section 3.2 has a thickness of 0.712 mm (±15µm). It should be mentioned that the crystal used in this section does not have a diamond substrate for efficient heat dissipation, as our goal is to understand thermal effects in the crystal. The electric field in the time domain for different duty cycles is represented in Fig. 3a where a 0.5 mm GaP crystal is used to detect the THz trace in the EOS setup. The electric fields are to scale with respect to each other but are offset in amplitude for better visibility. We obtain clean single-cycle THz pulses with minimal ringing, which results from water vapor absorption in the air. The increase in electric field amplitude for larger duty cycle, despite the constant pump pulse energy is a result of an integration effect in the photo detector: the number of the pulses per burst in higher duty cycles are more than at lower duty cycle, therefore more pulses get added up in the photo detector when the duty cycle is 50% rather than 10%. This however does not represent an increase in temporal or spectral dynamic range.

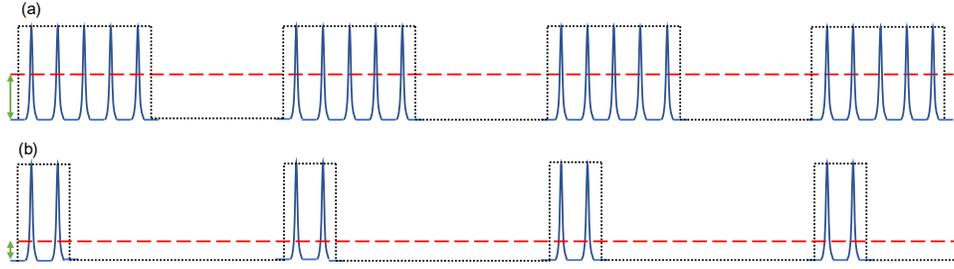

Fig. 2 Schematic representation of the pump pulse bursts at constant energy for different duty cycles. (a) chopper with 50% duty cycle. (b) chopper with 20% duty cycle. The red dashed line marks the average power level.

Using the emissivity value given in Table 1, we estimate the temperature of the crystal for different duty cycles at constant pulse energy as shown in Fig. 3b. It increases from 36°C to 66°C at 10% and 50% duty cycle, respectively. It shows about 40°C increase in the temperature by increasing the duty cycle, corresponding to an increase of the pump power.

In the next step, in order to investigate the generated THz power and conversion efficiency at fixed pulse energy and different pump power, the emitted THz power in each duty cycle is measured with the power sensor shown in Fig. 3c. The average power values show an increase with duty cycle, which is expected due to the increase in pump average power. For the same reason, the THz-to-optical conversion efficiency is expected to be constant since the pump pulse energy is not varied. However, the measurement shows a drop in efficiency of a factor of 1.75 for duty cycles >20%, indicating the influence of thermal effects. This effect can be seen, to a considerably lesser extent, in temperature data in Fig. 3b where the slope of the curve slightly increases for duty cycles >20%.

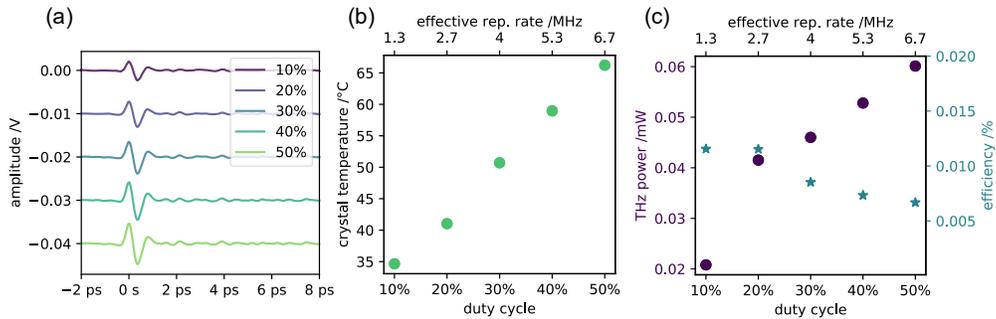

Fig. 3 (a) THz electric field sampled by a 0.5 mm GaP crystal in EOS setup for different duty cycles at constant pulse energy of 0.75 µJ on the crystal. (b) crystal temperature for different duty cycles at constant pulse energy, the top horizontal axis shows the effective repetition rate (c) left axis: measured THz average power using SLT power sensor, right axis: calculated optical-to-THz conversion efficiency.

In order to clarify that this saturation effect results only from the pump power and not the energy, we investigate the THz generation at constant power, but variable pulse energy in the next section.

## 3.2 THz generation efficiency vs. duty cycle at constant power on the crystal

In this section, the average pump power is kept constant and only the pulse energy is changed to investigate the influence of the pump pulse energy on the THz generation efficiency. In our setup, we can increase the average power, which at the same time, increases the pulse energy using the waveplate and TFP. By reducing the duty cycle of the pulse burst, we can keep the average power on the crystal constant and therefore only increase the pulse energy alone. This makes it possible to decouple effects resulting from an increase in average power (previous section) from those related to peak intensity/pulse energy. Fig. 4 illustrates this point and shows the pump pulse bursts at constant pump average power and different pulse energy for 50% and 20% of duty cycles. We use an average power of 1 W on the crystal for this experiment, i.e., 10 W of pump power before the optical chopper (operating at 10% duty cycle). For higher powers (corresponding to > 0.3 mJ/cm$^2$ fluence and > 4 kW/cm$^2$ average intensity), damage was consistently observed on the crystal. The EOS data is measured using the same 0.5 mm detection GaP crystal as in the previous section. The electric field in time domain for different duty cycles are shown in Fig. 5a. The observed increase in amplitude in the time traces by reducing the duty cycle represents an increase in THz power, which corresponds to an increase in pulse energy.

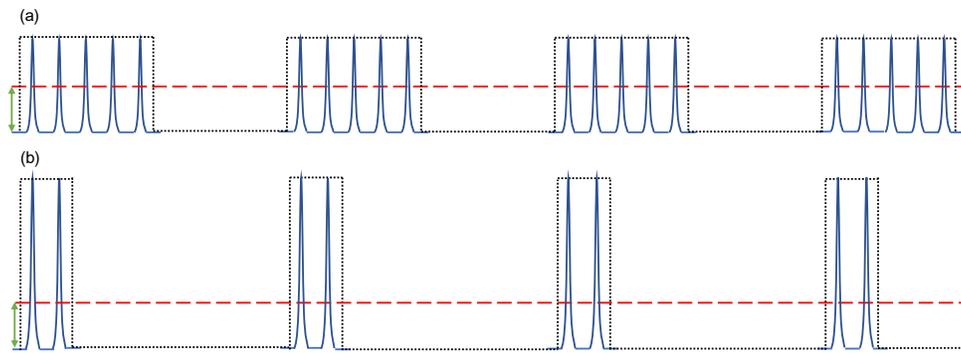

Fig. 4 The schematic representation of the pump pulse bursts in constant average power for different duty cycles. (a) 50% duty cycle of the chopper. (b) 20% duty cycle of the chopper. The red dashed line marks the average power level.

The thermal behavior of BNA at constant pump power is shown in Fig. 5b. Similar to section 3.1, the duty cycle and corresponding effective repetition rate are shown in the top and bottom horizontal axis, respectively. As shown in the plot, the measured temperature of the crystal changes only slightly between 63.6°C and 61.2°C. The almost constant temperature shows that the heating process is predominantly linear and does not strongly depend on the peak intensity. The observed slight increase in the temperature with higher pulse energy might be caused by the contribution of nonlinear effects, such as multi-photon absorption (MPA).

In order to study the effect of pulse energy variation on the emitted THz power and conversion efficiency, the detection crystal is replaced by the power sensor. Fig. 5c shows the measured THz power vs. duty cycle. As expected, the THz average power increases quadratically by decreasing the duty cycle which corresponds to higher pump pulse energy on the crystal (Fig 5c). This quadratic dependence on the duty cycle/pulse energy is expected for OR (which is a second order intra-pulse difference frequency generation process) in the low energy regime, similar to the GaP and Lithium Niobate results reported in [24]. This confirms that the saturation effect in section 3.1 is only due to the pump power and not the pulse energy, demonstrating that the main limiting effect in this high average power, high repetition rate regime are thermal effects. This points to future improvements of the setup by improving cooling even further, for example actively cooling it. This should also allow to apply the full pulse train, without the need to reduce the duty cycle of the laser. It is worth to highlight however that the exact physical mechanisms limiting the conversion efficiency and ending in damage for higher average powers require a significantly more detailed, separate investigation, which is out of the scope of this publication, but we plan to perform in a follow-up experiment.

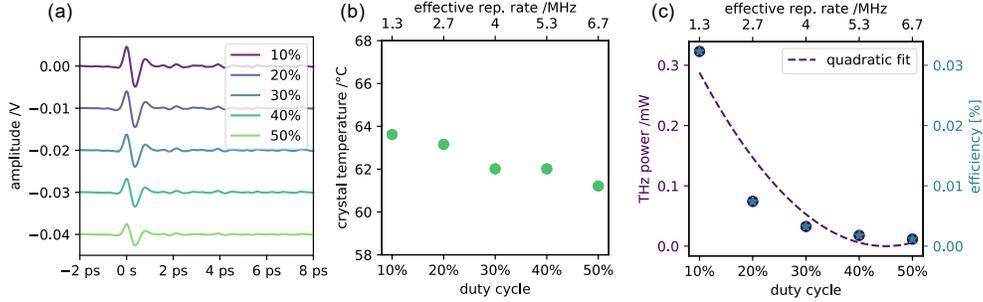

Fig. 5 (a) THz electric field sampled by a 0.5 mm GaP crystal in EOS setup for different duty cycles at constant pump power of 1 W on the crystal (b) crystal temperature for different duty cycles at constant pump power on the crystal, the top horizontal axis shows the effective repetition rate. (c) left axis: measured THz average power using SLT power sensor and linear quadratic fit (dashed line), right axis: calculated optical-to-THz conversion efficiency.

### 3.3 Highest THz average power with diamond-heatsinked BNA

As it has been shown in section 3.1, the highest conversion efficiency is achieved with the smallest duty cycle, since for this configuration, the pulse energy and the peak intensity are highest and thermal effects that decrease the conversion efficiency or lead to damage are reduced. In this section, for further power scaling, we use a constant duty cycle of 10% and increase the pump power from 0.3 W up to 2.9 W on the crystal. Moreover, to have better heat dissipation, a BNA crystal with thickness of 0.305 mm (±15 µm) glued on a diamond substrate is used. Since the pump beam first passes the diamond substrate to reach the crystal, a part of it gets reflected on the diamond surface. According to the Fresnel equations and by considering the refractive index of 2.4 for diamond at 1030 nm [25], a reflection coefficient of 0.17 is calculated for the diamond-air interface. Therefore, the corrected maximum pump power on the BNA is estimated to be 2.4 W, corresponding to a pulse fluence of 0.74 mJ/cm$^2$ where the 1/e$^2$ diameter of pump on the position of the crystal is 0.25 mm.

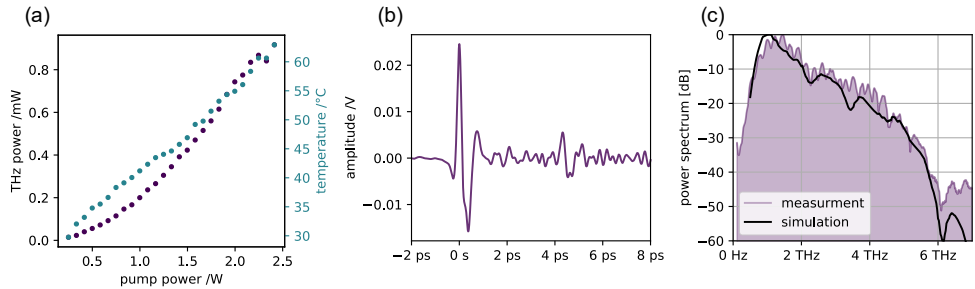

Fig. 6: Optimized configuration using diamond heatsinked sample: (a) left axis: THz power slope vs pump power at duty cycle of 10%, right axis: temperature of the crystal at each pumping power using infrared camera. (b) Measured EOS traces with 0.2 mm GaP as detection crystal. (c) corresponding spectrum (shaded purple) and simulated THz spectrum of expected bandwidth, power spectrum for a 85 fs pump pulse for BNA including the transmission function of the employed Teflon filter and low pass filtering of lock-in amplifier (black solid line).

Fig. 6a, left axis, shows the THz power vs. pump power, and temperature rise in this configuration. At pump powers lower than 2 W, the THz power increase is quadratic and for pump power above 2 W, it becomes linear. This can be a result of thermal effects which have been shown in section 3.1. At the maximum pumping power, we could reach a maximum THz

power of 0.95 mW which, to the best of our knowledge, is the highest power so far obtained with BNA. The conversion efficiency of 4×10$^{-4}$ without any irreversible damage on the crystal is achieved at this same maximum pump power. The advantage of the diamond heatsink can be clearly seen: the maximum temperature of the crystal with diamond substrate at 2.4 W incident power is 63°C whereas in the previous section a temperature of 66°C was reached already with 1 W of incident power. Better heat management due to the diamond substrate and a thinner crystal allowed us to pump this crystal higher than the BNA without heat sink used in section 3.1 and 3.2. Here we were able to use a maximum pump power of 2.4 W corresponding to an average intensity of 2500 kW/cm$^2$, compared to [21] with maximum intensity of 0.1 W/cm$^2$ at 10 Hz. We attribute this increase in pump power capability to the lower duty cycle of 10% as compared to the more commonly used 50% configuration as well as the diamond substrate, which enabled more efficient heat dissipation. For higher powers, which results in higher crystal temperature around 68°C, we observed thermal degradation and irreversible damage on the crystal.

The THz field at this maximum power is characterized by EOS using a 0.2 mm GaP detection crystal. The whole setup is purged with dry nitrogen to below 10% relative humidity to reduce water vapor absorption in the air. Fig. 6b shows the THz transient measured by EOS in the time domain under purged condition (10 averaged traces measured over 10 s). The corresponding power spectrum on a logarithmic scale is obtained by Fourier transformation from the measured waveform and is shown as shaded purple area in Fig. 6c. The spectrum is centered in the vicinity of 1.5 THz. The detected bandwidth spans up to 6 THz with a large dynamic range of >50 dB. In previously published work using BNA pumped with 800 nm [26] and with wavelengths between 1150 nm and 1500 nm [22,26] a significant dip in the THz spectrum was observed at 2 THz. This effect is greatly reduced in this work due to the different phase matching situation caused by different pumping wavelengths of 1030 nm which results in a broader, smooth THz spectrum from 0 to 6 THz.

In addition, to verify the generated spectral bandwidth in BNA, we numerically model the THz generation in BNA by solving the coupled wave equations in 1+1D, considering the temporal dimension and the propagation direction. The simulation is based on a split-step Fourier method and takes into account phase matching, pump depletion, and the nonlinear susceptibility of this material [27]. The refractive index and absorption coefficient of BNA in THz regime are taken from [28] and the nonlinear susceptibility is taken from [29]. Additionally, we take into account the transmission of the used Teflon filters [30], the low-pass filter of the Lock-in amplifier and the response function of the 0.2 mm GaP detection crystal calculated according to [31]. The black solid line in Fig. 6c represents the simulation result which shows excellent agreement with the measured spectrum.

## 4. Conclusion and outlook

We demonstrate OR in a diamond-heatsinked organic crystal BNA pumping by a nonlinearly compressed high-average power mode-locked thin-disk oscillator at MHz repetition rate, and present first details of thermal behavior of these crystals. In optimized conditions, we reach a maximum THz power of 0.95 mW with a smooth spectrum extending beyond 6 THz with a spectral dynamic range of more than 50 dB. To the best of our knowledge, this is the highest THz average power obtained using optical rectification in BNA. This source is very attractive for THz-TDS experiments where high power, broad bandwidth, high repetition rate and high dynamic range is advantageous. Furthermore, we present a first exploration of the thermal properties of these crystals under high-power pumping and shows that thermal effects represent the main limitation in the sense of thermal damage, and the temperatures below 60 °C are safe to apply in this repetition range regime.

In the near future, we plan to optimize our compression setup to sub-10 fs pulse duration which will enable significantly broader bandwidths with higher efficiency, since higher peak powers will be possible to apply without increasing the thermal load. Furthermore, for efficient heat management, actively cooling the crystal will allow us to scale the pump power further and reduce THz absorption. We believe this can lead this source well into the multi-10-mW regime. Moreover, it will allow us to continue our investigation of temperature-dependent effects in BNA.

**Acknowledgments**

We thank Tobias Buchmann for fruitful discussions during the first step of the experiments with organic crystals. This work was Funded by the Deutsche Forschungsgemeinschaft (DFG, German Research Foundation) – Project-ID 287022738 – TRR 196, and in part by the Alexander von Humboldt Stiftung (Sofja Kovalevskaja Preis).

**Disclosures**

The authors declare no conflicts of interest.

**Data Availability Statement**

Data underlying the results presented in this paper are not publicly available at this time but may be obtained from the authors upon reasonable request.